\documentclass[arxiv,reprint,superscriptaddress]{revtex4-1}
\usepackage[utf8]{inputenc}
\usepackage[english]{babel}
\usepackage{amsmath}
\usepackage{amsfonts}
\usepackage{amssymb}
\usepackage{graphicx}
\usepackage{epstopdf}
\usepackage{url}
\begin{document}
\author{P.N.~Skirdkov}
\author{K.A.~Zvezdin}
\author{A.D.~Belanovsky}
\affiliation{A.M.~Prokhorov General Physics Institute, Russian Academy of Sciences, Vavilova 38, 119991 Moscow, Russia and Moscow Institute of Physics and Technology, Institutskiy per.~9, 141700 Dolgoprudny, Russia}
\author{J.~Grollier}
\author{V.~Cros}
\affiliation{Unité Mixte de Physique CNRS/Thales and Université Paris-Sud 11, 1 ave A. Fresnel, 91767 Palaiseau, France}
\author{C.A.~Ross}
\affiliation{Massachusetts Institute of Technology, Cambridge, MA 02139, USA}
\author{A.K.~Zvezdin}
\affiliation{A.M.~Prokhorov General Physics Institute, Russian Academy of Sciences, Vavilova 38, 119991 Moscow, Russia and Moscow Institute of Physics and Technology, Institutskiy per.~9, 141700 Dolgoprudny, Russia}
\title{Domain wall displacement by remote spin-current injection}
\begin{abstract}
We demonstrate numerically the ability to displace a magnetic domain wall by a remote spin current injection. We consider a long and narrow magnetic nanostripe with a single domain wall (DW). The spin-polarized current is injected perpendicularly to the plane of the film (CPP) through a small nanocontact which is located at certain distance from the domain wall initial position. We show theoretically that the DW motion can be initiated not only by conventional spin-transfer torque but also by indirect spin-torque, created by a remote spin-current injection and then transferred to the DW by the exchange-spring mechanism. An analytical description of this effect is proposed. This finding may lead to a solution of bottleneck problems of DW motion-based spintronic and neuromorphic devices with perpendicular spin-current injection.
\end{abstract}
\maketitle
The study of domain wall (DW) dynamics in magnetic nanostripes has attracted much attention in the last decade due to both fundamental \cite{Saitoh-2004} and applied \cite{Parkin-2008} motivations. On the one hand complex collective magnetization dynamics can be induced by several means, on the other hand DW-based nanostructures are very promising for creating magnetic logic and memory devices \cite{Allwood-2005, Hayashi-2008, Behin-Aein-2010}. Initially, it was proposed to control DW dynamics by magnetic fields \cite{Ono-1999, Atkinson-2003, Beach-2005}. However, this approach is hardly suitable for close-packed arrays of nanoscale devices due to significant cross-talk effects. An alternative is to use current-induced domain wall motion that has been subject of many experimental \cite{Grollier-2003, Yamaguchi-2004, Klaui-2005, Hayashi-2007, Ravelosona-2007} and theoretical \cite{Bazaliy-1998, Tatara-2004, Li-2004, Zhang-2004, Thiaville-2005, Khvalkovskiy-2013} studies. The interest to the current-induced DW dynamics is significantly encouraged by the developments of promising magnetic-based neuromorphic devices \cite{Sharad-2012}, spintronic logics \cite{Behin-Aein-2010, Currivan-2012}, race-track memory \cite{Parkin-2008} and spintronic memristors \cite{Wang-2009, Chanthbouala-2011, Locatelli-2014}.
\par
\begin{figure}[h!]
\centering
\includegraphics[width=0.45\textwidth]{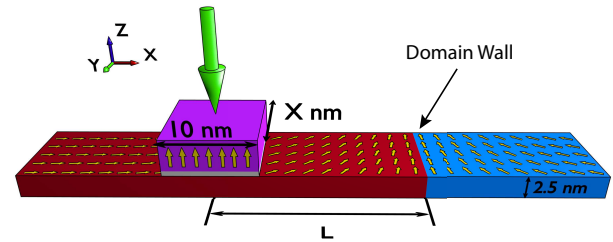}
\caption{(Color online) The studied system, composed by a permalloy nanostrip with size $3000\times X\times2.5~nm^3$ and a nanocontact with size $10\times X~nm^2$. \label{fig:devive}}
\end{figure}

The studied nanostructures are usually composed of a long and narrow magnetic film (nanostrip) containing the domain wall. For this geometry, there are two possible directions of the current: current-in-plane (CIP), when the spin polarized current is flowing in the plane of the magnetic film, and current perpendicular to the plane (CPP), when the spin polarized current is flowing perpendicular to the magnetic film. Recent theoretical \cite{Khvalkovskiy-2009} and experimental \cite{Boone-2010, Metaxas-2013} studies show that in the CPP configuration the DW velocities can be up to two orders of magnitude larger then in the CIP configuration for equal current densities. Thus the CPP configuration requires relatively low current densities for efficient DW dynamics excitation \cite{Chanthbouala-2011}. The drawback of such configuration however is very high electric currents required for efficient DW motion, since a direct current action on the DW is required to initiate its motion \cite{Khvalkovskiy-2009}. Moreover, in the conventional geometry of the neuromorphic logic devices \cite{Sharad-2012} the input current contacts and the DW are separated by a distance $L\gg\Delta$, where $\Delta$ is the typical DW width, and the direct current action is simply impossible. The question of a possible non-contact (indirect) interaction between the CPP current, localized in the contact, and the DW remains unresolved. Recently an all-magnonic mechanism of DW displacement has been proposed \cite{Yan-2011}, in which the DW dynamics is induced by the spin-waves excited remotely from the DW initial localization. To achieve relatively high DW velocities using this mechanism, however, one needs to excite high-amplitude magnons using very high magnetic fields \cite{Han-2009, Jamali-2010}, which are hardly achievable in real-life applications. Here we propose the study of the DW motion induced by a remotely localized CPP spin-current injection that can help to solve these issues. We investigate numerically the DW motion for the case when the spin current is flowing perpendicular to the plane through a small localized nanocontact (see Fig.~\ref{fig:devive}), which is located at a certain distance from the initial DW localization. We show theoretically that DW displacements of several hundred nanometers can be obtained by a very low remotely injected CPP spin-polarized current (about $50~\mu A$), in contrast to the conventional case when the current flows through the entire DW. 
\begin{figure*}[t]
\centering
\includegraphics[width=\textwidth]{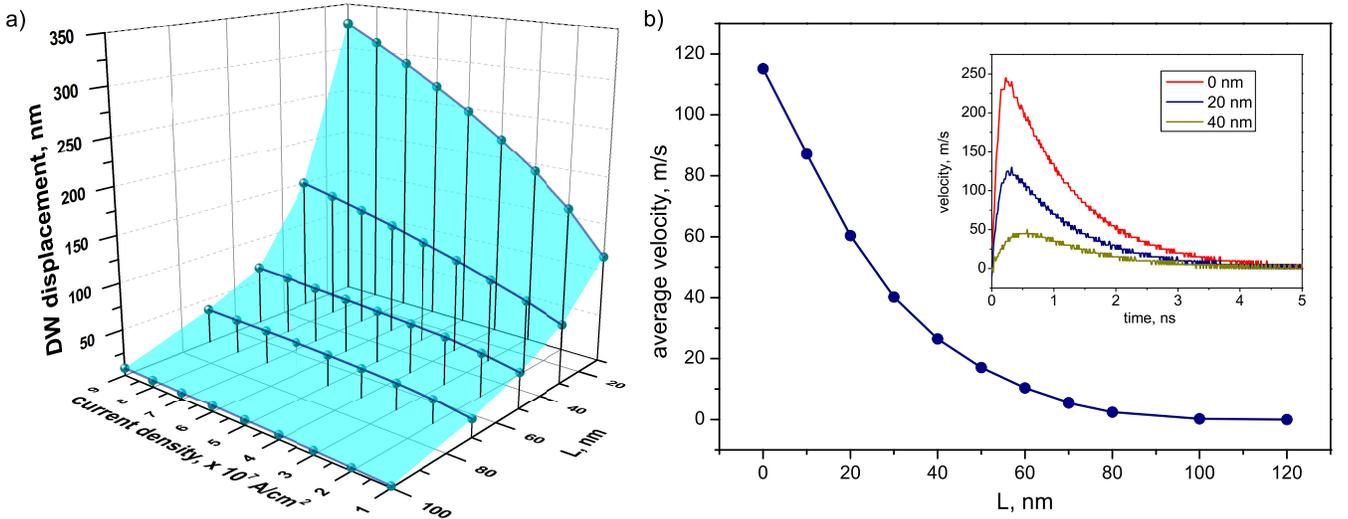}
\caption{(Color online) a) DW displacement for different initial distances $L$ between the injection contact and the DW and for different current densities in case of the nanostrip width $X=10~nm$. b) Dependence of the average velocity on the distance $L$. The speed averaging is carried out for the first $3$ nanoseconds, since during this time the DW has already been shifted almost the maximum possible distance. In the box: time dependence of the velocity for $L=0,20,40~nm$. The results are presented for the current density $J=5\times10^7~A/cm^2$ and the nanostrip width $X=10~nm$. \label{fig:1}}
\end{figure*}
In addition to the evident practical interest, this system is of a high fundamental importance too, because the mechanism of interaction between remotely localized current and the domain wall is not obvious. We demonstrate here that the perpendicular to the plane electric current localized in a small nanocontact generates an in-plane charge-less spin current in the nanowire, and that this spin current may effectively excite the remote DW's motion. An analytical description of this effect based on the soliton perturbation theory was proposed.
\par
The magnetization dynamics in the nanostrip is described by the Landau-Lifshitz-Gilbert (LLG) equation with an additional term responsible for the spin transfer \cite{Slonczewski-1996, Berger-1996}:
\begin{equation}
\label{eq:LLG}
\dot{\mathbf{M}}=-\gamma\mathbf{M}\times\mathbf{H}_{eff}+\mathbf{T}_{STT}+\frac{\alpha}{M_S}(\mathbf{M}\times\dot{\mathbf{M}}),
\end{equation}
where $\mathbf{M}$ is the magnetization vector, $\gamma$ is the gyromagnetic ratio, $\alpha$ is the Gilbert damping constant, $M_S$ is the saturation magnetization and $H_{eff}$ is the effective field consisting of the magnetostatic field, the exchange field and the anisotropy field. The spin transfer torque $\mathbf{T}_{STT}$ is represented by two components \cite{Zhang-2002, Xia-2002}: a Slonczewski torque (ST)
$\mathbf{T}_{ST}=-\gamma\frac{a_j}{M_S}\mathbf{M}\times(\mathbf{M}\times\mathbf{m}_{ref})$
and a field-like torque (FLT)
$\mathbf{T}_{FLT}=-\gamma b_j(\mathbf{M}\times\mathbf{m}_{ref})$
where $\mathbf{m}_{ref}$ is a unit vector along the magnetization direction of the reference layer. The ST amplitude is given by $a_j=\hbar JP/2deM_S$, where $J$ is the current density, $P$ is the spin polarization of the current, $d$ is the thickness of the free layer and $e>0$ is the charge of the electron. The amplitude of the FLT is given by $b_j=\xi_{CPP}a_j$, where $\xi_{CPP}$ can be larger than $0.4$ in case of an asymmetric magnetic tunnel junction. \cite{Chanthbouala-2011}
\par
The studied system is composed of a permalloy $Ni_{81}Fe_{19}$ (Py) nanostrip magnetized in-plane and containing a head-to-head domain wall, and a reference nanocontact with fixed out-of-plane magnetization (see Fig.~\ref{fig:devive}). Hence, we consider magnetization dynamics in the free layer only, and the nanocontact acts as a static spin polarizer. The size of nanostrip has been choosen to be $3000\times X\times2.5~nm^3$, and the size of the nanocontact was $10\times X~nm^2$, where $X=10$ to $110~nm$. The stripe dimensions have been chosen large enough in order to have a negligeable influence of the edges on the main features of the DW motion. The Py magnetic parameters used in the modelling are: $M_S=800~emu/cm^3$, the exchange constant $A=1.3\times10^{-6}~erg/cm$, $\alpha=0.01$, spin polarization $P=0.31$, and the bulk anisotropy is neglected. The CPP current has been switched on in the nanocontact at the time $t=0$. To investigate the remote influence of the CPP spin-polarized current on the domain wall, we have performed a series of simulations using our micromagnetic finite-difference code SpinPM \cite{SpinPM} based on the fourth-order Runge-Kutta method with an adaptive timestep control for the time integration and a mesh size $2\times2~nm^2$. In order to focus on the spin torque mechanisms of the DW dynamics excitation both Oersted field and thermal fluctuations have not been taken into account. However, it is worth noting that for the studied geometry the amplitude of Oersted field even under the contact is not more than 15 Oe, and at the distance about 10 -- 15 nm it almost completely disappears. Hence, it leads only to a change of the DW initial position by 10 -- 15 nm. It is also should be noted, that although here we presented the results for the head-to-head DW, the results for tail-to-tail DW are the same, except the direction of DW motion is reversed.
\par
\begin{figure*}[t!]
\centering
\includegraphics[width=\textwidth]{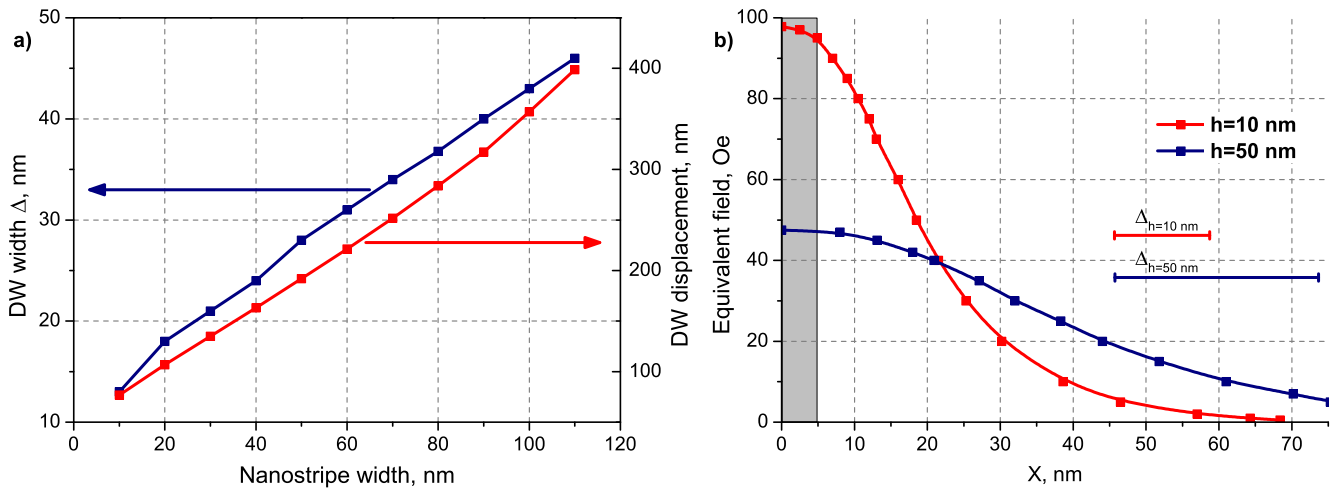}
\caption{(Color online) a) Dependence of DW width $\Delta$ and DW displacement on the nanostripe width. Initial distance is $L=40~nm$, current density is $J=5\times10^7~A/cm^2$. The blue and red arrows indicate to which axis the curves belong. b) Dependence of the equivalent magnetic field on X position for two nanostripe thicknesses. The grey region represents the current nanocontact. The blue and red segments show the width of the DW for the cases of $h=10$ and $h=50$. \label{fig:2}}
\end{figure*}
The displacements of the DW for different distances $L$ and current densities are presented in Fig.~\ref{fig:1}(a). The evolution of the DW velocity after switching on the CPP spin-polarized current (at t=0) for different initial distances between the center of the contact and the DW is demonstrated in the inset of Fig.~\ref{fig:1}(b). The dynamics of the DW is as follows: once the CPP spin-polarized current is switched on in the nanocontact, after some small delay period, the DW starts to accelerate for about $0.5$ ns, and decelerates until a complete stop after a few nanoseconds.
\par
To explain these observations, let us consider the magnetization in the current injection region. Since the strip is thin enough (in comparison with its length), at the initial time the magnetization in the strip beneath the contact is oriented strictly along the strip. The DW width $\Delta$ is obtained by fitting of micromagnetic data using a traveling wave ansatz $\theta(\delta x)=2\arctan(\exp[\delta x/\Delta])$. In our case DW width $\Delta\approx 13$ to $46~nm$ depending on the nanostripe width (see Fig.~\ref{fig:2}(a)). It is to be emphasized that $\Delta$ must be considerably less than the distance $L$, at which the nanocontact is separated from the DW. Therefore there is no direct action of the electric current on the DW. However, the presence of the DW even at a considerable distance leads to a small tilt of the magnetization, in other words, to appearance of the perturbed region ("tail"). The action of the spin-transfer on the DW can be decomposed into two steps. At first, under the influence of the current flowing through the contact, the spins that are beneath the injection contact experience a torque, which leads to the local increase of the Y component of magnetization in the contact region (about $30~emu/cm^3$ for our set of parameters). The fact that a local pulse of the spin-polarized current can influence a DW remotely is determined by the presence of exchange stiffness in the magnetic structure of the material (in a soft medium such an effect is obviously absent). Then, through the exchange-spring mechanism, this disturbance is transmitted from the DW "tail" located inside the nanocontact region directly to the domain wall, thereby causing the DW drift. Previously exchange-springs were studied in heterophase systems \cite{Kneller-1991, Zeng-2002}, but similar effects can be observed in homogeneous systems as well. Indeed, in our case, each subsequent magnetic moment of the DW "tail" is deflected at a slight angle from  the direction of the previous one. If the external action (CPP local current in our case) deflects one or more of the magnetic moments slightly (edge of the spring is deformed), then due to the strong exchange interaction the subsequent magnetic moments are also deflected one by one. As a result, the exchange-spring is straightened, pushing the DW. After the complete straightening of the exchange-spring, the DW stops accelerating and starts to slow down due to the damping and after some time finally stops. Hence the effect is defined by the static exchange-spring tension. The numerical simulations demonstrate a delay in the onset of the DW motion with respect to the time of the current switching, corresponding to the propagation time of the spring excitation (static tension). It is worth stressing that in this case of indirect action of the spin transfer, the angular momentum is transferred to the DW not by conduction electrons, but by the charge-less spin current, due to the exchange-spring interaction. Moreover when the nanostrip width is increased, the domain wall becomes wider, therefore the "tail" of the DW also becomes longer, the DW is affected by the current by a longer time, and consequently the DW should be displaced over larger distances. This effect is perfectly confirmed by the simulations (see Fig.~\ref{fig:2}(a)).
\par
To estimate the magnitude of the exchange-spring interaction, we obtained the dependence of the equivalent field (magnetic field with direction opposite to the DW motion direction, which has to be applied to counterbalance the action of the exchange-spring static tension) on the X coordinates (see Fig.~\ref{fig:2}(b)). We see that, even at the distance of several times greater than the DW width $\Delta$ the equivalent field is still large enough to displace the DW significantly. It also shows that at large thicknesses of the nanostrips the equivalent field decays more slowly. Despite the fact that for large thicknesses the equivalent field near the nanocontact is smaller, the effectiveness of the exchange-spring will still be higher for large thicknesses than for smaller ones, as follows from results of the DW displacement (see Fig.~\ref{fig:2}(a)). The reason for this is that with increasing thickness of the nanostrip, the efficiency of the magnetic field is growing faster than the efficiency of the local current contact. As a result, although the actual magnetic field required to balance the action of the exchange-spring proves to be smaller, the force acting on the DW will be larger for larger thicknesses.
\par
For the final test of the proposed mechanism, the simulations have been performed for the case of $\alpha=1$. Such a large damping eliminates the effect of spin waves, as they fade out before reaching the DW. Also in this case there is practically no movement by inertia (as soon as the external forces stop acting, the DW should immediately stop its free motion). However, the result of our simulations shows that the DW is still displaced by a distance of about $60~nm$ (with a width of the DW $\Delta=13~nm$), which corresponds to the distance at which the equivalent  field almost becomes zero. From this we can conclude that the spin waves do not determine the effect, which is caused only by the exchange-spring static tension. Another important result is that the considered mechanism of DW dynamics excitation (via static tension of the exchange-spring) does not require an alternating current or magnetic field in contrast to the case in which the DW is excited by spin waves \cite{Han-2009, Jamali-2010}. This makes it promising for practical application, like racetrack memory, magnetic logic and neuromorphic devices.
\par
For an analytical insight into this mechanism let us write Eq.\eqref{eq:LLG} in spherical coordinates with the energy represented by $E=A((\nabla\theta)^2+\sin^2\theta(\nabla\phi)^2)+2\pi M^2_S\cos^2\theta-K_\perp\sin^2\theta\cos^2\phi$, where $A$ is the exchange constant, $K_\perp$ is anisotropic constant, $\phi$ is the polar angle and $\theta$ is the azimuth angle:
\begin{align}
\label{eq:sp1}
\begin{split}
-\sin\theta\dot{\theta}-\alpha\sin^2&\theta\dot{\phi}=\frac{1}{2}\omega_\perp\sin^2\theta\sin 2\phi- \\ & -\frac{2\gamma A}{M_S}\sin^2\theta\phi''+\gamma a_j(x,t)\sin^2\theta \end{split} \\ \label{eq:sp2}
\begin{split}
\sin\theta\dot{\phi}-\alpha\dot{\theta}&=-\frac{2\gamma A}{M_S}\theta''+\sin\theta\cos\theta\cdot \\ & \cdot(\frac{2\gamma A}{M_S}(\phi')^2-\omega_\parallel-\omega_\perp\cos^2\phi ) \end{split}
\end{align}
where $\omega_\perp=2\gamma K_\perp/M_S$, $\omega_\parallel=4\pi\gamma M_S$ and $a_j(x,t)$ is not equal to zero only in the contact region. Due to the shape anisotropy the magnetization lies almost entirely in the plane. The simulation shows that the deviation of the magnetization from the plane is not more than $3~\%$ of the $M_S$. With this in mind, consider a small deviation in $\theta$: $\theta=\pi/2+\theta_1$ ($\theta_1<<1$). In this case, Eq.\eqref{eq:sp1},\eqref{eq:sp2} take the following form:
\begin{align}
\label{eq:1}
& -\dot{\theta_1}-\alpha\dot{\phi}=\frac{\omega_\perp}{2}\sin 2\phi-\frac{2\gamma A}{M_S}\phi''+\gamma a_j(x,t) \\ \label{eq:2}
& \dot{\phi}-\alpha\dot{\theta_1}=-\frac{2\gamma A}{M_S}\theta_1''-\theta_1(\frac{2\gamma A}{M_S}(\phi')^2-\omega_\parallel-\omega_\perp\cos^2\phi)
\end{align}
Taking into account that $\omega_\perp/\omega_\parallel\ll 1$ and $2\gamma A/M_S l^2\ll\omega_\parallel$, where $l$ is the typical spatial scale, and neglecting small quantities, we can rewrite Eq.\eqref{eq:2} in following form: $\theta_1=\dot{\phi}/\omega_\parallel$. Substituting this result into the Eq.\eqref{eq:1}, we obtain:
\begin{equation}
\label{eq:sin-Gor}
\ddot{\phi}-c^2\frac{\partial^2 \phi}{\partial x^2}+\omega^2_0\sin\phi\cos\phi=-\alpha\omega_\parallel\dot{\phi}-\gamma\omega_\parallel a_j(x,t)
\end{equation}
where $c^2=8\pi\gamma^2 A$ and $\omega^2_0=\omega_\parallel\omega_\perp$. Eq.\eqref{eq:sin-Gor} is the modified version of the sine-Gordon equation. The solution of this equation with zero right-hand side of the equation is represented by a kink soliton propagating with constant velocity $\tan(\tilde{\phi}_0/2)=\exp\left\{\pm(x-vt-x_0)/\Delta\right\}$, where $v$ is the velocity of the domain wall, $x_0$ is the initial distance between the DW and nanocontact center, $\Delta=\Delta_0\sqrt{1-v^2/c^2}$ and $\Delta_0=c/\omega_0$. Since the maximum velocity in modelling is about $250~m/s$ and $c\approx 1000~m/s$, we can estimate $1\ge\sqrt{1-v^2/c^2}\ge 0.96$, and therefore $\Delta\approx\Delta_0$.
\par
The micromagnetic modelling demonstrates acceleration and deceleration of the DW. To take this into account let us assume that the right-hand side of the equation only slightly modifies the DW's profile but changes the velocity. This assumption was confirmed by simulations. In this case we can use the soliton perturbation theory. We represent the solution of the Eq.\eqref{eq:sin-Gor} as $\phi=\phi_0+\phi_1$, where $\phi_1\ll 1$ and $\phi_0$ is the modified kink $\tan(\phi_0/2)=\exp\left\{\pm(x-x_c(t)-x_0)/\Delta_0\right\}$ with $x_c(t)$ is treated as the DW's position. Then after linearisation and neglecting small values Eq.\eqref{eq:sin-Gor} assumes the form:
\begin{equation}
\label{eq:lin_sin-Gor}
\begin{split}
& (\frac{\partial^2}{\partial t^2}-c^2\frac{\partial^2}{\partial x^2}+\omega^2_0\cos 2\phi_0)\phi_1= \\ & = -\alpha\omega_\parallel\dot{\phi_0}-\gamma\omega_\parallel a_j(x,t)+\frac{\ddot{x}_c}{\Delta_0}\sin{\phi_0}
\end{split}
\end{equation}
Let us define the operator $\hat{L}=\partial^2/\partial t^2-c^2\partial^2/\partial x^2+\omega^2_0\cos 2\phi_0$ and function $f(\phi_0)=-\alpha\omega_\parallel\dot{\phi_0}-\gamma\omega_\parallel a_j(x,t)+(\ddot{x}_c/\Delta_0)\sin{\phi_0}$. Using this notation Eq.\eqref{eq:lin_sin-Gor} can be written as $\hat{L}\phi_1=f(\phi_0)$. According to the Fredholm alternative \cite{scott}, this equation has a solution if and only when right-hand side $f(\phi_0)$ of the equation is orthogonal to the eigenfunction of operator $\hat{L}$ with zero eigenvalue, which can be found from equation $\hat{L}\phi_1^{(0)}=0$. For the present problem the required eigenfunction takes the form $\phi_1^{(0)}=\partial\phi_0/\partial x$. Then, considering that $\dot{\phi_0}=\mp(\dot{x}_c/\Delta_0)\sin\phi_0$ and $\partial\phi_0/\partial x=\pm\sin\phi_0/\Delta_0$, the solvability condition is:
\begin{equation}
\label{eq:int}
\left(\alpha\omega_\parallel\frac{\dot{x}_c}{\Delta^2_0}+\frac{\ddot{x}_c}{\Delta^2_0}\right)\langle\sin^2\phi_0\rangle-\frac{\gamma\omega_\parallel}{\Delta_0}\langle a_j(x,t)\sin\phi_0\rangle=0
\end{equation}
where $\langle...\rangle$ means integration over $x$. After integration, taking into account that $a_j(x,t)$ is not equal to zero only inside the contact region, we obtain a Newton-like equation of motion:
\begin{equation}
\label{eq:sol}
m\ddot{x}_c=-\alpha\omega_\parallel m\dot{x}_c+F(x_c)
\end{equation}
where $m=1/(\pi\gamma^2\Delta_0)$ is the effective mass of the DW, $F(x_c)=(a_j\omega_\parallel/\pi\gamma)\cdot\arctan\left[\exp\left\{(d/2-L-x_c)/\Delta_0\right\}\right]-\arctan\left[\exp\left\{(-d/2-L-x_c)/\Delta_0\right\}\right]$ is the force created by the current and $d$ is the size of the contact along the X axis.
\par
\begin{figure}[t!]
\centering
\includegraphics[width=0.45\textwidth]{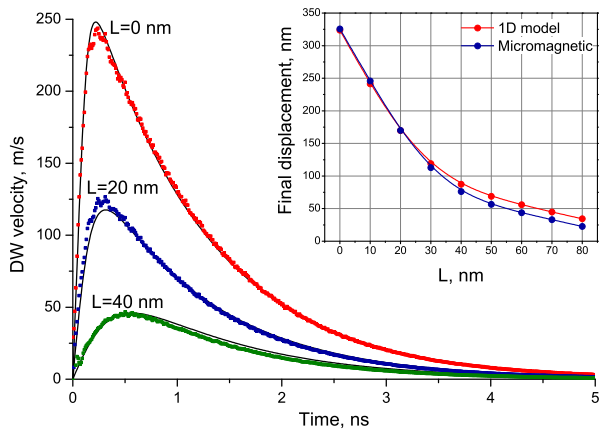}
\caption{(Color online) Time dependence of the DW's velocity obtained analytically (continuous black line) for different distances $L$ in comparison with the micromagnetic modelling results (points). In the inset: comparison of the final DW's displacements obtained using 1D analitic model and by micromagnetic simulations as a function on $L$. \label{fig:teor}}
\end{figure}

Time dependence of the velocity of the domain wall for different $L$, obtained from Eq.\eqref{eq:sol}, in comparison with the micromagnetic modelling results is shown in Fig.~\ref{fig:teor}. As can be seen from the graph, the proposed analytical model demonstrates a good agreement with the results of micromagnetic simulation. It should be noted that although for small $L$ values theory predicts an absolutely correct final DW displacement, with increasing $L$ the difference between simulations and theoretical predictions slightly increases, up to $10~nm$ for $L=100~nm$ (see inset in Fig.~\ref{fig:teor}).
The reason for this discrepancy lies in the fact that the analytical 1D model is based on the rigid soliton model. Under this assumption we neglect DW deformation and consider that the the action of the current contact propagates instantaneously. But in the case of large $L$ values the exchange-spring needs some time, corresponding to the propagation time of the spring excitation, to start pushing the DW. During this time the effective in-plane spin current slightly attenuates. Because of this the 1D model slightly overestimates the final displacement of the DW for larger $L$ values.
\par
In conclusion, we have demonstrated theoretically the possibility of spin current induced domain wall motion in the CPP geometry, when the DW is initially located outside the nanocontact region. Although velocities in this case are lower than in the usual CPP case (about $500~m/s$) \cite{Metaxas-2013}, they are still higher than the velocities in the CIP geometry; the required currents are very low (about $50~\mu A$), in contrast to the case when the current flows through the entire sample \cite{Khvalkovskiy-2009,Chanthbouala-2011,Metaxas-2013}. We have shown that the DW dynamics in this case is induced by indirect spin-torque, created by a remote spin-current injection, which is transferred then to the DW by the exchange-spring mechanism. The analytical description of this effect based on the soliton perturbation theory was proposed. Although this mechanism of DW dynamics excitation can be used by itself, it can also be effectively used to depin a DW, when magnetization dynamics is driven by less effective methods (e.g. in-plane current injection). On this basis, the remotely localized contact injection of CPP spin-polarized current becomes a very promising option for practical applications, such as racetrack memory, magnetic logic and neuromorphic devices.
\par
Financial support by the European Research Council (Starting Independent Researcher Grant No. ERC 2010 Stg 259068), the French ANR grant ESPERADO 11-BS10-008, RFBR Grants No. 12-02-01187 and No. 14-02-31781, CNRS PICS Russie No. 5743 2011 and Skolkovo Inst. of Technology.
\par
\providecommand{\noopsort}[1]{}\providecommand{\singleletter}[1]{#1}%
\end{document}